\documentclass[usenatbib]{mnras}
\usepackage[T1]{fontenc}
\usepackage{aecompl}
\usepackage{amsmath,amsfonts,amssymb}
\usepackage[pdftex]{graphicx}
\usepackage{aas_macros}
\usepackage[usenames]{xcolor}
\usepackage{color}
\usepackage{textcomp,gensymb}
\usepackage{times}
\usepackage{subfig}
\usepackage{stfloats}
\usepackage[utf8]{inputenc}
\newcommand{\newc}[1]{{#1}}
\title[The circumbinary disc of AB~Aur]{Binary-induced spiral arms inside the disc cavity of AB~Aurigae}

\author[Poblete et al.]{\parbox{\textwidth}{Pedro P. Poblete$^{1,2,3}$, Josh Calcino$^{4}$, Nicol\'as Cuello$^{5,1,6}$, Enrique Mac\'ias$^{7,8}$,\\ \'Alvaro Ribas$^{8}$, Daniel J. Price$^{9}$, Jorge Cuadra$^{1,2,10}$ and Christophe Pinte$^{9,5}$ }\vspace{0.15cm}\\
$^{1}$Instituto de Astrof\'isica, Pontificia Universidad Cat\'olica de Chile, Santiago, Chile,\\
$^{2}$N\'ucleo Milenio de Formaci\'on Planetaria (NPF), Chile, \\
$^{3}$Astrophysikalisches Institut, Friedrich-Schiller-Universität Jena, Schillergäßchen 2–3, 07745 Jena, Germany\\
$^{4}$School of Mathematics and Physics, The University of Queensland, QLD 4072, Australia \\
$^{5}$Univ. Grenoble Alpes, CNRS, IPAG, F-38000 Grenoble, France, \\
$^{6}$Univ Lyon, Univ Lyon1, Ens de Lyon, CNRS, Centre de Recherche Astrophysique de Lyon UMR5574, F-69230, Saint-Genis-Laval, France,\\
$^{7}$Joint ALMA Observatory, Alonso de C\'ordova 3107, Vitacura, Casilla 19001, Santiago, Chile  \\
$^{8}$European Southern Observatory (ESO), Alonso de C\'ordova 3107, Vitacura, Casilla 19001, Santiago, Chile \\
$^{9}$School of Physics \& Astronomy, Monash University, VIC 3800, Australia.\\
$^{10}$Departamento de Ciencias, Facultad de Artes Liberales,
Universidad Adolfo Ib\'a\~nez, Avenida Padre Hurtado 750, Vi\~na del Mar, Chile.}

\begin{document}
\date{Accepted ... Received ...}

\pagerange{\pageref{firstpage}--\pageref{lastpage}} \pubyear{2020}

\maketitle

\label{firstpage}

\begin{abstract}
In this work we demonstrate that the inner spiral structure observed in AB~Aurigae can be created by a binary star orbiting inside the dust cavity. We find that a companion with a mass-ratio of $0.25$, semi-major axis of 40~au, eccentricity of 0.5, and \newc{inclination of $90\degree$ produces gaseous spirals closely matching the ones observed in $^{12}$CO (2-1) line emission. Based on dust dynamics in circumbinary discs \citep{Poblete+2019}, we constrain the inclination of the binary with respect to the circumbinary disc to range between $60\degree$ and $90\degree$.} We predict that the stellar companion is located roughly $0.18\arcsec$ from the central star towards the east-southeast, above the plane of the disc. Should this companion be detected in the near future, our model indicates that it should be moving away from the primary star at a rate of 6 mas/yr on the plane of the sky. Since our companion is inclined, we also predict that the spiral structure will appear to change with time, and not simply co-rotate with the companion.
\end{abstract}


%%%%%%%%%%%%%%%%%%%%%%%%%%%%%%%%%%%%%%%%%%%%%%%%

\section{Introduction}
\label{sec:intro}

Transition discs (TDs) are a class of protoplanetary discs identified by their central cavities depleted in dust grains \citep{Strom+1989}. Originally thought to be an evolutionary stage that all protoplanetary discs pass through, it is now thought that TDs actually arise due to companion disc interactions \citep{Marsh+1992,Marsh+1993,Owen+2016}. Therefore TDs may be ideal targets to constrain planet formation theories. 
Spatially resolved observations in the infrared and at millimetre wavelengths have revealed a complex morphology in many TDs. The observation of asymmetric dust horseshoes further suggest a dynamical origin for the central cavity as these structures have been interpreted as vortices generated by planet-disc interactions \citep[e.g. see][]{vanderMarel+2013, Perez+2014, Pacheco-Vazquez+2016, Fuente+2017}, gas over-densities orbiting a circumbinary disc \citep{Ragusa+2017, Price+2018, Calcino+2019, Poblete+2019}, or pile-ups of dust at the apocentre of an eccentric disc \citep{Ataiee+2013}. Spiral arms observed in scattered light also suggest a dynamical origin for TD cavities, since spirals are expected to arise as companions excite Lindblad resonances in the disc \citep{Goldreich+1979,Goldreich+1980}. Although evidence is mounting that the structures observed in TDs arise due to companion-disc interactions, it is still not clear what the properties of these companions are. Such companions could be planetary \citep{Quillen+2005,Dong+2015a}, stellar binary \citep{Ragusa+2017, Price+2018, Calcino+2019, Poblete+2019} or stellar flyby \citep{Clarke&Pringle1993,Pfalzner2003,Cuello+2019b} in nature. 

The young star AB~Aurigae \citep[estimated age of $4\pm1$ Myr][hereafter AB~Aur]{DeWarf+2003} is a spectacular and puzzling example of a TD. With an A0 spectral type and mass $2.4 \pm 0.2\ M_{\odot}$, AB~Aur is one of the closest Herbig Ae stars located at 162.9 $\pm$ 1.5 pc \citep{Gaia2018}. The central star is surrounded by a small 2-5 au radius disc detected in the near-infrared \citep{Millan-Gabet+2006,diFolco+2009} and an outer disc which begins at roughly 70-100 au from the central source \citep{Pietu+2005}, extending to roughly 450 au \citep{Henning+1998}. The outer disc around AB~Aur also shows several prominent features such as a large cavity \citep{Pietu+2005,Hashimoto+2011,Tang+2012}, spiral arms \citep{Fukagawa+2004,Corder+2005,Hashimoto+2011,Tang+2017}, and dusty clumps \citep{Tang+2012,Pacheco-Vazquez+2016,Fuente+2017,Tang+2017}, among others. 

Inside the cavity, two prominent spiral-like features have been detected in $^{12}$CO (2-1) emission with high-resolution observations \citep{Tang+2017}. The spiral features are four times brighter than the surrounding medium, and seem to be co-located with asymmetrical structures seen in near-infrared observations \citep{Hashimoto+2011}. The inner regions coincident with the spiral arms appear misaligned with respect to the outer disc by $\sim20 \degree$ \citep{Hashimoto+2011, Riviere-Marichalar+2019}. Additional spiral arms outside the cavity have also been detected, and seem to be concentrated in a specific azimuthal section of the disc \citep{Fukagawa+2004,Hashimoto+2011}. 

It has been suggested that the spiral arms in AB~Aur are the result of either one \citep{Dong+2016}, or multiple companions connected to the spiral arms \citep{Tang+2017}, both internal and external to the spiral arms. It is also proposed that this companion can excite the Rossby Wave Instability to trigger vortex formation \citep{Fuente+2017}, producing the dust asymmetry observed at millimetre wavelengths. However, dust asymmetries can also be generated in circumbinary discs \citep{Ragusa+2017, Price+2018, Calcino+2019, Poblete+2019}. Additionally, the gaseous spirals within the cavity could be signatures of binary-disc interactions.

In this paper, we demonstrate that the spiral arms observed inside the cavity of AB~Aur can be explained by the presence of an inclined and eccentric inner binary. This adds new evidence to the binary hypothesis made in \citet{Poblete+2019} based on the mm-dust distribution. We describe the numerical methods and initial conditions in Section~\ref{sec:methods}. We report our results in Section~\ref{sec:results}. We discuss them in Section~\ref{sec:discussion} and draw our conclusions in Section~\ref{sec:conclusion}.

%%-------------------------

\section{Methods}
\label{sec:methods}

\subsection{SPH simulations}
We perform 3D hydrodynamical simulations of circumbinary discs (CBDs) using the {\sc Phantom} smoothed particle hydrodynamics (SPH) code \citep{PricePH+2018b}. We use the same setup as the `e50–i90' case in \citet{Poblete+2019} for both disc and binary, which we describe briefly below. 

We model both stars as sink particles \citep{Bate+1995} with an accretion radius of 1 au. The binary is initialised with a mass ratio of $q=M_2/M_1=0.25$ with $M_1=2\,M_{\odot}$ and $M_2=0.5\,M_{\odot}$. The companion is placed on an orbit with a semi-major axis of $40$~au, and an eccentricity $e_{\rm B}=0.5$, giving a period of $\sim$160 yrs. The binary is inclined by $i_{\rm B}=90\degree$ with respect to the circumbinary disc mid-plane. The initial argument of periapsis ($\omega$) and the line of nodes ($\Omega$) are both set to $0\degree$. The change and implications of these values will be discussed in Section~\ref{sec:binparams}.

The gas disc in our simulation is initialised with $5\times 10^6$ SPH particles in Keplerian rotation with a total gas mass of $0.01\ M_{\odot}$. The surface density profile is given by a power law, $\Sigma \propto R^{-1}$. \newc{The disc is vertically isothermal} and the temperature profile follows a shallower power law, $T\propto R^{-0.3}$, giving a scale height of $H/R = 0.06$ at $R_{\rm in}$ and $H/R = 0.1$ at $R_{\rm out}$. We set the SPH viscosity parameter $\alpha_{\rm AV}\approx0.3$, which gives a mean \citet{Shakura&Sunyaev73} disc viscosity of $\alpha_{\rm SS}\approx5\cdot10^{-3}$ \citep[c.f.][]{Lodato&Price2010}. We did not include dust grains in our simulation.

\subsection{Radiative Transfer Calculations}\label{sec:RT_model}

We create synthetic observations of the output from our SPH simulation using the Monte Carlo radiative transfer code {\sc mcfost} \citep{Pinte+2006,Pinte+2009}. We then simulate the ALMA synthetic observations using the \texttt{simutil} tools in {\sc casa} (v. 5.5.0) and the same $uv$-coverage as the observations. Finally, images were obtained using the same procedure as with the observations (see Section \ref{sec:co_obs}).

Since we do not include dust grains in our simulation, we construct a dust model using the gas distribution in our SPH simulation and assuming a dust population with a grain-size distribution $dn/ds \propto s^{-m}$ between $s_{\min} = 0.01$ $\mu$m to $s_{\max}=1$ mm with $m=3.5$. This gives an opacity of $\kappa = 15\ $cm$^2$g$^{-1}$ at a wavelength of 1 mm. The total dust mass is calculated assuming \newc{the typical gas-to-dust ratio value of 100}, where the gas mass is taken directly from our simulation.

The dust grain opacities are temperature independent and calculated assuming spherical and homogeneous grains. The temperature of the central star is set to 10,000 K \citep{Ancker+1998} and the luminosity is matched to the estimated value of 52 $L_\odot$ by \newc{setting the radius to 2.5 $R_\odot$} and assuming it emits as a blackbody. Radiation from the secondary star is also included, where we assume a temperature of 3800 K and radius of 1.32~$R_\odot$, giving a blackbody luminosity of approximately 0.19~$L_\odot$ \citep{Siess+2000}. The disc is passively heated and we assume that the dust and gas are in thermal equilibrium. We assume an abundance ratio of $^{12}$CO-to-H$_2$ of $10^{-4}$ when computing the $^{12}$CO (2-1) line emission intensity.

We used $10^8$ Monte Carlo photon packets to compute the temperature and specific intensities. Images were then produced by ray-tracing the computed source function. We assume an inclination of $i = 26^{\circ}$, a position angle PA$=-36^{\circ}$ \citep{Tang+2017, Riviere-Marichalar+2019}, and a source distance of 163 pc \citep{Gaia2018}.

\subsection{Observational Data}\label{sec:co_obs}

To confront our simulations with observations, we retrieved the data used in \citet{Tang+2017} from the ALMA archive (project 2015.1.00889.S). These observations were taken in band 6, and include $^{12}$CO (2-1) and continuum data at 1.3\,mm with spatial resolution of $\sim$0.05" \citep[see][for a full description of the observations]{Tang+2017}. We performed three rounds of phase self-calibration using the continuum observations, which improved the continuum peak SNR from 38 to 76. After applying the self-calibration solutions, we imaged the $^{12}$CO (2-1) emission using natural weighting, a $uv$-tapering of 0.08", and removing baselines shorter than 160\,k$\lambda$, since their poor sampling produced substantial artefacts. The resulting cube has a restoring beam of 0.12"x0.09" and an RMS of 3.5\,mJy/beam for a 150\,m/s channel width. Finally, the corresponding moments 0 and 1 were produced using the \emph{bettermoments} software \citep{bettermoments}, using $\sigma$-clipping at the 3\,rms level. A similar process was used to create the corresponding moments for the simulations.

%%-------------------------

\section{Results}
\label{sec:results}

\begin{figure*}
\centering
\begin{center}
    {\includegraphics[width=1.\textwidth]{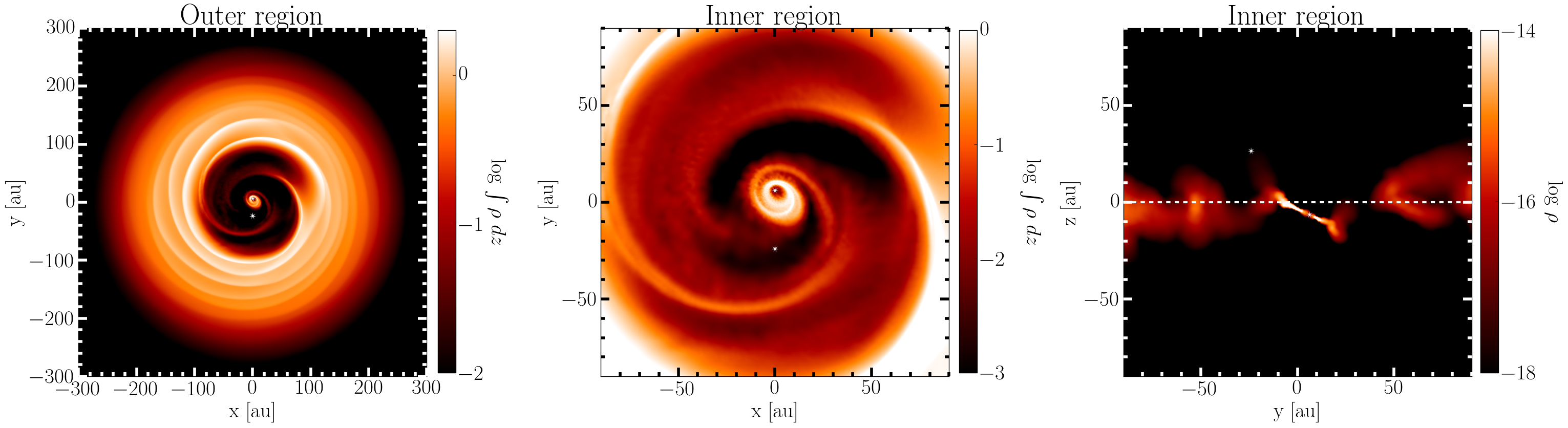}}%
    \caption{Gas morphology for the simulation once it reaches a quasi-steady state at 30 binary orbits. The left panel shows the entire gas disc, which highlights the structures of the outer region ($r>90$ au), while the \newc{middle and right panel show a zoom-in at the inner region ($r<90$ au). In the first two panels, the disc is seen face-on while the binary, located inside the cavity, is perpendicular to the disc. The third panel shows a cross-section at $x=0$ au on the $zy$-plane. The dashed line represents the mid-plane of the circumbinary disc.}} 
    \label{fig:simulation}
\end{center}
\end{figure*}

Figure \ref{fig:simulation} shows the gas surface density after approximately 30 binary orbits. We divide the disc in two sectors, the outer (the left panel when $r>90$ au), and the inner region (\newc{the middle and} right panels when $r<90$ au), for easier discussion of the spiral features propagating through the circumbinary disc and inside the cavity. The cavity is contained in the inner region; it is the transition between the circumbinary disc at the outer region and the circumprimary disc.

\subsection{\newc{Spiral Arms and Circumprimary Disc}}
\label{sec:spirals}

The outer region exhibits multiple spirals which are spiral density waves generated by the periodic gravitational interaction between the binary and the inner rim of the gas disc. Such a pattern is similar to that observed in HD~142527 \citep{Avenhaus+2014}, a well studied binary system. The densest spiral arms seem to concentrate in the upper-half of the disc, which matches the location of the binary pericenter, and they remain there throughout the simulation (i.e. long after 30 binary orbits). As \citet{Poblete+2019} describe, this region favours the formation of dust clumps, which rotate along the inner edge of the disc. Spiral arms outside the cavity have been seen in AB~Aur in near-infrared observations \citep{Fukagawa+2004, Hashimoto+2011}.

The inner region of the circumbinary discs displays two prominent spiral arms, similar to the observed spirals in the binary systems [BHB2007]~11 \citep{Alves+2019}, FS~Tau A \citep{Yang+2020}\newc{, and GG~Tau A \citep{Phuong+2020}. Due to the inclination of the companion, these spiral arms are shifted out of the plane with respect to the circumbinary disc.} There are also two spiral arms protruding off of the circumprimary disc, which may be primary and secondary inner wakes of the companion star. Such spirals are well studied for massive planetary bodies on circular, co-planar orbits \citep{Dong+2015a}, and co-rotate with the companion. However as our companion is on an eccentric and inclined orbit, we expect these wakes to have a more complicated behaviour. The two prominent spirals seem to connect the inner edge of the gas disc with the innermost regions, producing a flow of material from the outer region. Therefore, they enrich the central part, feeding the circumprimary disc, and central \newc{stars}.

\newc{We note that the circumprimary disc is i) inclined with respect to the circumbinary disc and ii) eccentric (see Figures~\ref{fig:simulation} and \ref{fig:spiral_evolution}). This is due to the binary misalignment and eccentricity, which cause complex three-dimensional morphologies inside the cavity. Such an inclined circumstellar disc can experience Kozai-Lidov oscillations due to perturbations from the outer companion \citep{martin2014, zanazzi2017}. Therefore, we do not expect that the circumprimary disc will remain on a fixed orientation with respect to the outer disc and companion. This mechanism is important to consider when inferring the dynamics or stability of misaligned inner discs in observed systems.}

\subsection{Orbital Dependence of the Inner Structures}
\label{sec:orbital_dependence}

\begin{figure*}%
\begin{center}
    {\includegraphics[width=0.85\textwidth]{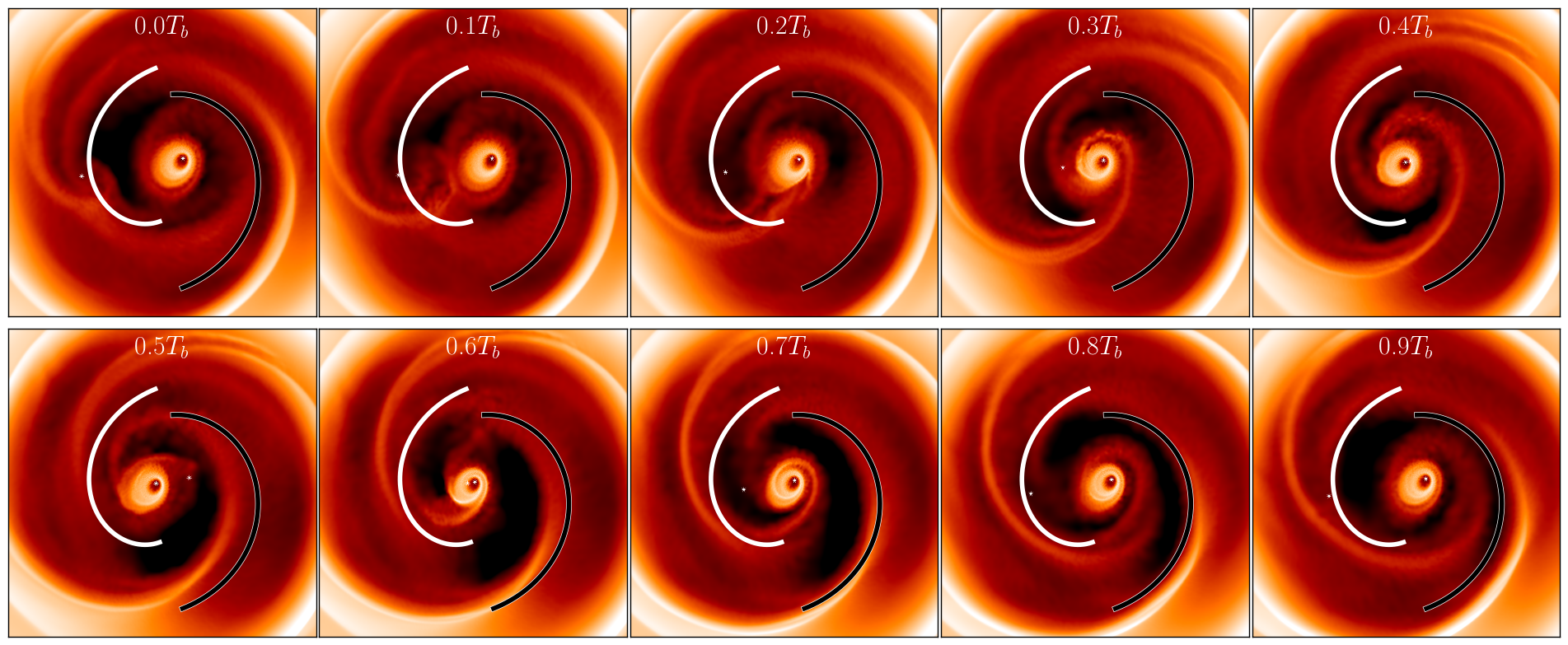}}%
    \caption{Inner gas region of the simulation for a whole orbital period ($T_\mathrm{b}$). The orbit is divided into 10 frames to show the changes of gas spirals in the cavity due to the presence of the binary. The white and black lines represent the eastern [Equation \eqref{eq:eastern_spiral}], and the western [Equation \eqref{eq:western_spiral}] spiral respectively. The rotation of the system disc-binary is done according to the best match described in Section \ref{sec:CO_maps}. Since our companion is inclined with respect to the disc, it is below the plane of the disc for $0\ T_\mathrm{b} \leq 0.5\ T_\mathrm{b}$, and above the plane of the disc for $0.5\ T_\mathrm{b} \leq 1\ T_\mathrm{b}$. Our closest agreement to the fitted spirals from \citet{Tang+2017} occurs at $\sim 0.7~T_\mathrm{b}$, when the companion is above the plane of the disc, and close to the primary star along the line of sight.}% \djp{colour bar should use max 3 orders of magnitude, even better use linear scale, like obs}}%
    \label{fig:spiral_evolution}%
\end{center}
\end{figure*}
\begin{figure*}%
\begin{center}

    {\includegraphics[width=0.77\textwidth]{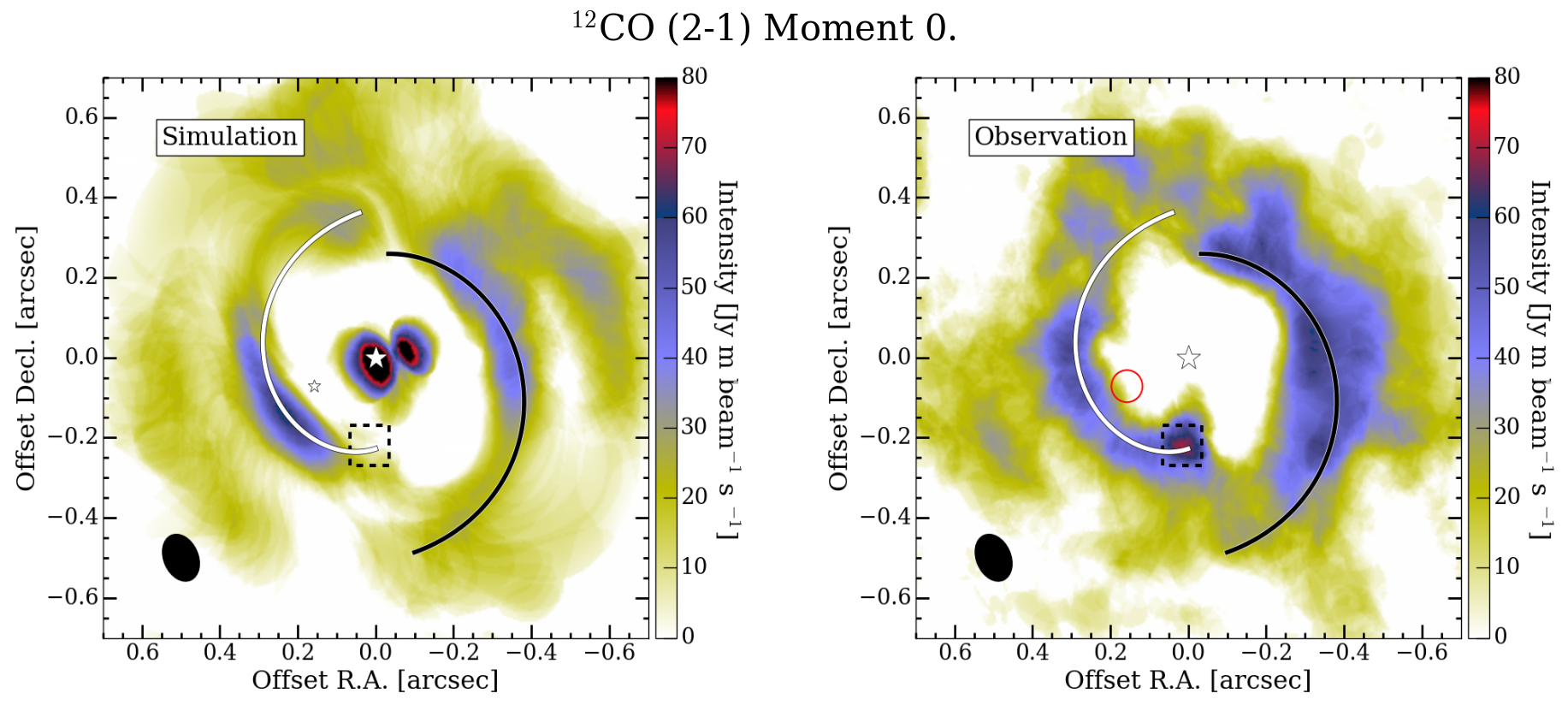}}\\
    {\includegraphics[width=0.77\textwidth]{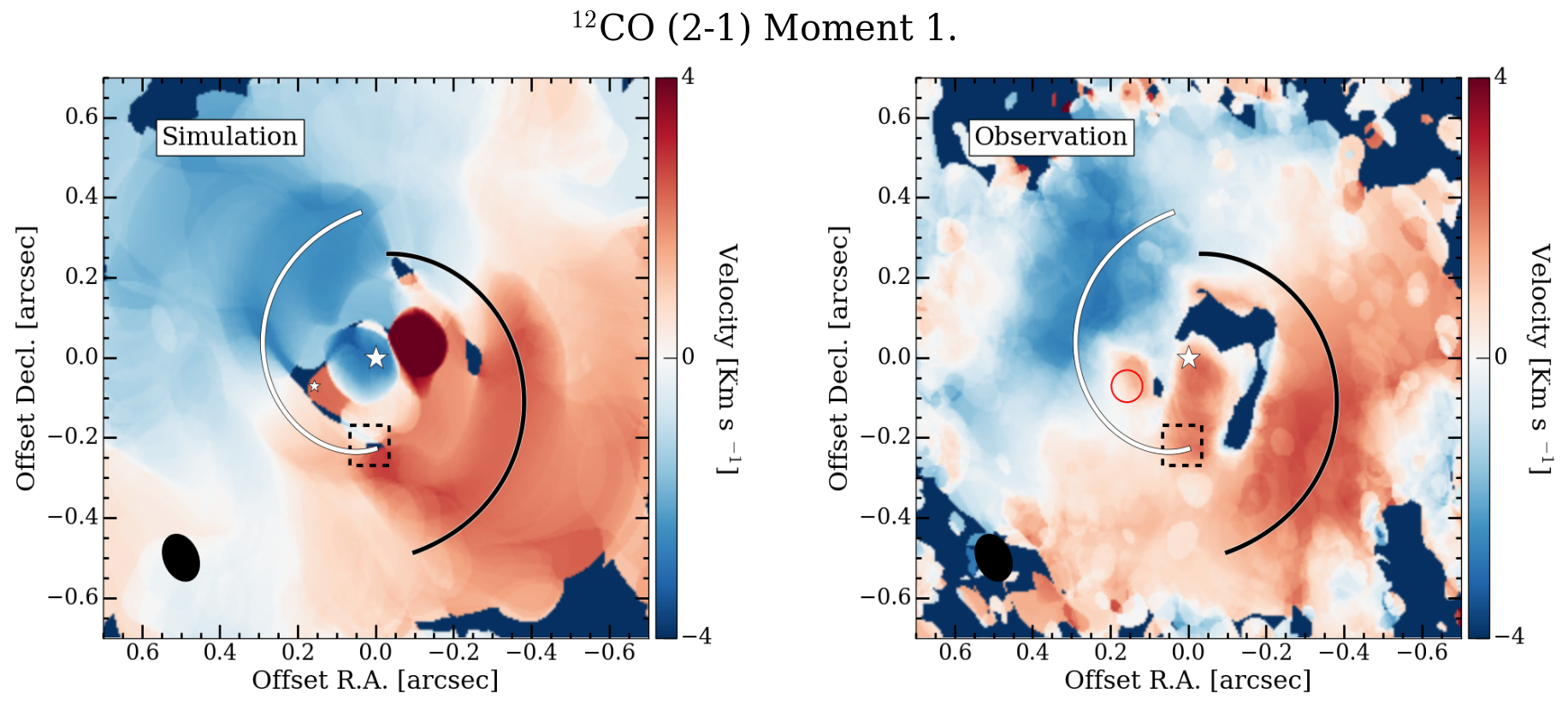}}%
    \caption{Comparison of $^{12}$CO (2-1) moments between our simulation (left panels), and the observation (right panels). The moment 0 is displayed in the first row and the moment 1 in the second row. The spiral functions are represented as the white line (eastern arm) and the black line (western arm). The central star is shown at the center of each panel. In our simulations, we mark the location of the secondary star as a smaller star mark; however, we mark as a red circle the predicted site that it would have the companion in the observations. We highlight the observed hot-spot in the observation as a dashed square in all panels. The black ellipse at the bottom-left corner of each panel represents the beam size employed.}
    \label{fig:moment_0}%
\end{center}
\end{figure*}

In order to study the spiral structures inside the cavity in more detail, in Figure \ref{fig:spiral_evolution} we plot the surface density for ten time-steps of the binary orbit after the simulation has reached a quasi-steady state. It is clearly seen that the spiral pattern and the gas distribution change markedly over the binary orbit, in strong contrast to what occurs for companions on co-planar and circular orbits \citep{Dong+2015a}. Therefore we expect that the spiral structure in AB Aur will change over time, and not simply co-rotate with a companion in the cavity.

The apparent chaotic environment of the inner region can be explained by the gravitational effects of the secondary star with the surrounding gas. When the secondary is close to periastron ($t \sim\ 0.4~T_{\mathrm b}$), the two spirals fall to the circumprimary disc. When the secondary star is at the apoastron ($t \sim\ 0.0~T_{\mathrm b}$), one of the spirals starts to feed it, instead of falling toward the central region. The spiral morphology inside the cavity therefore appears to be a complicated mix of streamers feeding the central circumprimary disc, and spiral density waves propagated from the companion. The orbital dependence of these inner structures, especially the two main spiral arms, could be used as a predictor of the secondary star location. This aspect will be discussed in more detail in Section \ref{sec:discussion}. 

\subsection{Integrated CO Emission}
\label{sec:CO_maps}

\cite{Tang+2017} observed two prominent spirals in the inner regions of AB~Aur in $^{12}$CO (2-1) emission. The authors provide the following analytical functions to each spiral (as logarithmic spirals) 
\begin{align}
r(\theta) &= 0.85 \cdot  e^{-21 \cdot \theta}\ \mathrm{arcsec}, \label{eq:eastern_spiral}\\%0.85\cdot e^{-21 \theta}\\
r(\theta) &= 0.38 \cdot  e^{-12.5 \cdot \theta}\ \mathrm{arcsec},  \label{eq:western_spiral}%\cdot  e^{-12.5 \theta}
\end{align}
where $\theta$ is the angular coordinate in radians. Equation \eqref{eq:eastern_spiral} represents the eastern spiral, and Equation \eqref{eq:western_spiral} the western spiral. We use these functions as a reference to rotate our simulation when conducting the radiative transfer modelling. Rotating our model such that the eccentricity vector of the binary companion points at a $\textrm{PA} \sim 100 \degree$ provides a close match to the spiral structure. We have plotted the rotated simulation along to the analytical spirals, namely, the spirals inside the cavity of AB~Aur observed with $^{12}$CO (2-1) line emission from \cite{Tang+2017} (see Figure \ref{fig:spiral_evolution}).

Figure \ref{fig:moment_0} shows the comparison between the integrated $^{12}$CO (2-1) line emission of our simulation computed when $t \sim 0.7\ T_b$ (upper-left panel), and the integrated $^{12}$CO (2-1) line emission observed in AB~Aur (upper-right panel). The spiral functions are plotted in both panels, white for the eastern arm, and black for the western arm. Our model matches relatively well the location of the spiral arms, and apart from the hot-spot at the south side, the intensity also matches. \newc{The model shows an excess of CO emission near the primary star with respect to the observation. However, we note that our simulations do not include mechanisms such as magneto-hydrodynamical effects, stellar winds, or photoevaporation, all of which could decrease the amount of gas in the inner regions \citep{Suzuki+2010,Alexander+2014}. Additionally, dust radial drift could reduce the gas-to-dust ratio in the circumprimary disc \citep{Nakagawa+1986}, thus increasing its optical depth and yielding lower CO emission (see Appendix~\ref{A1} to watch the effects of a lower gas-to-dust ratio). Regardless, these processes are beyond the scope of this paper, and they should not impact our main results.}

\subsection{CO Kinematics}
\label{sec:vel_maps}

We show the comparison between the CO velocity maps of our simulation and observations of AB~Aur in the bottom row of Figure~\ref{fig:moment_0}. The kinematics inside the cavity of AB~Aur appear to be strongly perturbed from what would be expected for an inclined disc in Keplerian rotation. This has also been confirmed from HCO$^+$ observations presented in \citet{Riviere-Marichalar+2019}, where the inner regions in particular ($r < 0.6$ arcsec) show twisted iso-velocity lines. It has been suggested by several studies on AB~Aur that the inner cavity is misaligned with respect to the outer disc \citep{Hashimoto+2011,Tang+2017,Riviere-Marichalar+2019}, which can explain the peculiar inner kinematics. This is in agreement with our model, as material inside the cavity becomes misaligned with respect to the outer disc due to the orbit of the companion \newc{(see right panel of Figure \ref{fig:simulation})}. 

\begin{figure}
\centering
\begin{center}
    {\includegraphics[width=0.48\textwidth]{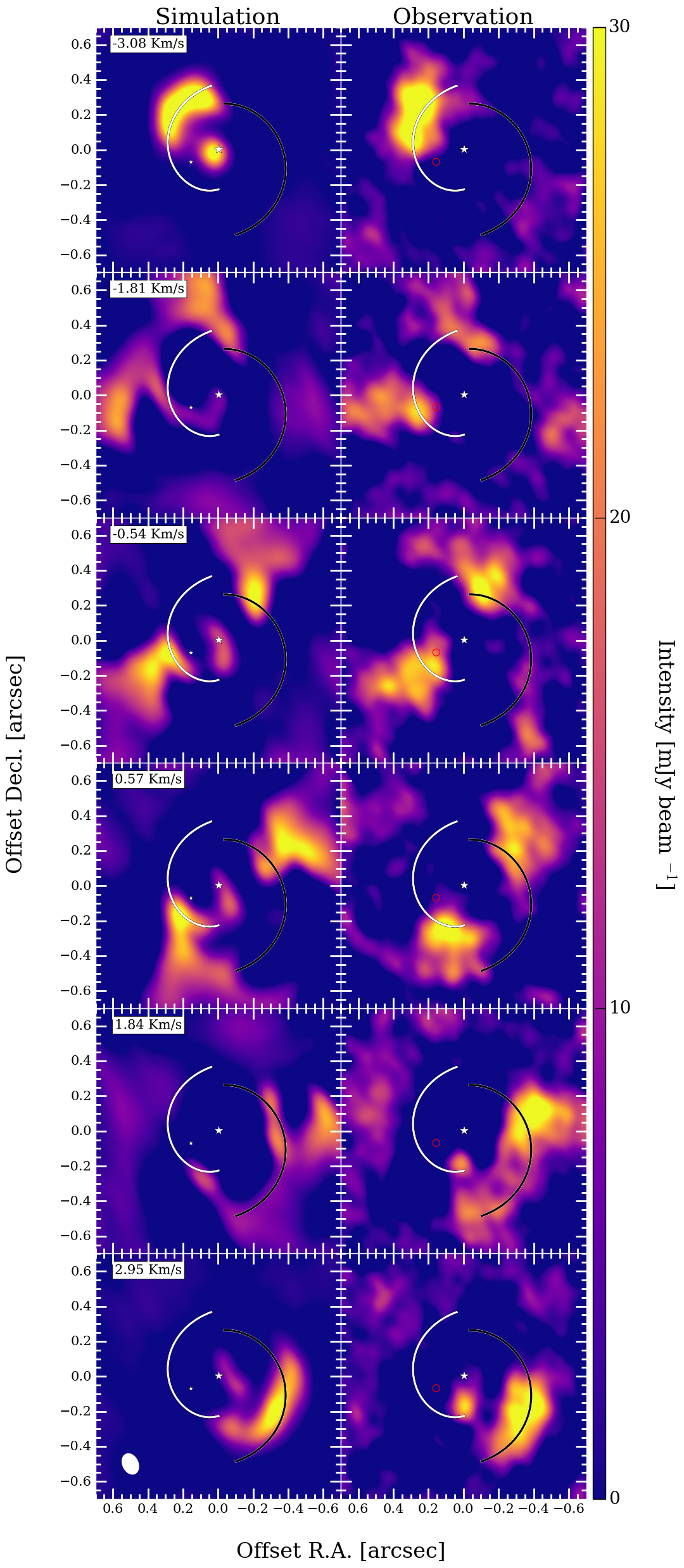}}%
    \caption{Comparison between $^{12}$CO (2-1) channel maps of our simulation (left column), and reprocessed observations (right column). We assume a systemic velocity of $v_{\rm sys}=5.85$ Km/s. We use the same marks in Figure~\ref{fig:moment_0} to show the stars, and the same prescription for the spiral arms. The beam employed in each case is displayed at the last row of the first column as a white ellipse at the bottom-left corner.}
    \label{fig:chan_map}
\end{center}
\end{figure}

 Figure \ref{fig:chan_map} compares a selection of the channel maps generated from our simulation with our reprocessed channel maps of AB~Aur. In the channel maps of AB~Aur, we first note that the CO emission appears to trace the spiral structure, which is consistent with our model. The propagation angle of the iso-velocity curves differs slightly between our model and the observations, which may be a result of the inner regions of our disc being at a different inclination angle than the inner cavity region of AB~Aur. 

The emission tracing the \newc{southern tip of the} eastern spiral continues in the high velocity channels \newc{(i.e. for $v_\textrm{los} > 1.84$ km/s)}, which is also reproduced in our model. \cite{Tang+2017} suggest this may be the signature of a circumplanetary disc, since it is also co-located with a hot-spot in the integrated CO emission (see Figure \ref{fig:moment_0}). In our model, this high velocity material is not directly connected with the companion \newc{because the companion is above the circumbinary disc plane. This corresponds instead to the material that is falling onto the circumprimary disc, which is characterised by a higher line-of-sight velocity compared with the rest of the disc. This effect is due to the binary inclination with respect to the outer disc. This complex flow appears clearly in Figure~\ref{fig:spiral_evolution} for $t \sim 0.7~T_{\mathrm{b}}$.} This material is not as bright in our model compared with the observations due to it being shielded by the circumprimary disc and therefore having a lower temperature than the eastern spiral, which is not being shielded. In AB~Aur, the hot-spot may be accretion flows onto the circumprimary disc which are being directly illuminated by the primary star, explaining the excess emission, and its high velocity.

There are also more tentative features in the channel maps worth discussing. In our simulation, the iso-velocity curves of the channel maps do not propagate outward smoothly as would be expected from an unperturbed Keplerian flow. This is particularly noticeable in our channels with a low line-of-sight velocity, such as when $v_\textrm{los} = [-0.54, 0.57]$ km/s. These features in our model are not a product of the sparse $uv$-coverage, as they are also evident when we assume perfect $uv$-coverage. Turning our attention to the observations of AB~Aur in the right column of Figure \ref{fig:chan_map}, there may be some analogous structures in the iso-velocity curves. However we caution that these structure could be artefacts. 

Localised perturbations in the iso-velocity curves have been suggested to indicate the presence of planetary mass companions co-located with the perturbations \citep{Perez+2018,Pinte+2018,Pinte+2019}. However when the companion is of stellar mass the perturbations are no longer localised, but are evident across much of the disc. Similar perturbations in the iso-velocity curves were reported in \citet{Calcino+2019} for IRS 48. Our simulations indicate that we should also expect to see similar structures in AB~Aur. Confirming these perturbations with higher spectral and spatial resolution observations (with more complete $uv$-coverage) would a good test of our binary hypothesis.

\section{Discussion}
\label{sec:discussion}

Thus far, no successful detections of an inner companion in AB~Aur have been made. Hints for the presence of a companion inside the cavity can be found from spectro-astrometric and NIR interferometric observations \citep{Baines+2006}, and radio-emissions \citep{Rodriguez+2007}. Nevertheless, these observations can be interpreted differently, and they do not provide strong evidence for a companion in AB~Aur. 

 The evidence presented here, and the study of dusty clumps in \citet{Poblete+2019}, support a stellar binary hypothesis for AB~Aur, which explains a considerable number of observed features. The dependence of the inner structure on the orbital phase of the companion star (see Section~\ref{sec:orbital_dependence}) provides a way to predict the companion location. Our best match to the spirals seen in CO line emission (Figure~\ref{fig:moment_0}) is reached when the binary companion is above the plane of the disc and moving away from the primary star towards the east-southeast. Previous searches for a companion in this disc may have missed the companion since it was much closer to the line of sight of the primary star (see panel $0.6\,T_\mathrm{b}$ of Figure~\ref{fig:spiral_evolution}). It is then understandable how such a massive companion may have been missed by previous observations of this system \citep{Pirzkal+1997}. In our model the companion is currently located at roughly $r \sim 0.18\arcsec$ from the central source, with a $\rm PA\sim 100\degree$ and is moving away from the primary star at a rate of roughly 6 mas/yr. These quantities are model-dependent; however, we expect that changing the binary parameters could also result in a good match to the observed spiral structure with the companion at a different location in the cavity. This aspect is discussed further below.

\subsection{Binary Orbital Parameters}
\label{sec:binparams}

Our simulation provides a compelling match to the spiral structure in AB~Aur. However, once again we caution that other binary configurations may produce a similar double spiral morphology with the companion at a different location in the cavity. The possible parameter space is large, and changing the eccentricity, inclination, semi-major axis, mass-ratio, the argument of periapsis, could all change the spiral morphology. In this work we have not conducted an exhaustive search of this parameter space to rule out other possible configurations. Other works have explored \newc{ranges} of binary mass-ratio, eccentricity, and semi-major axis \citep[e.g. see][]{Thun+2017}. However, the parameter space including inclination and argument of periapsis has not been explored as extensively. A limited portion of this parameter space was explored in \citet{Price+2018} for HD~142527 (see their figures~1 \newc{and 2}).

It is worth mentioning \newc{the main} differences in disc morphology that occur when the binary plane is misaligned with respect to the circumbinary disc. Firstly, when the binary is co-planar, the circumbinary disc almost always becomes eccentric \citep{Ragusa+2017, Thun+2017, Calcino+2019}, with the primary star near one foci of the ellipse. Another effect on an eccentric disc is the precession of the cavity shape \citep{Dunhill+2015}. This would change the structures inside the cavity. However when the binary plane is misaligned, the circumbinary disc is not strongly eccentric in general \citep[e.g. see Figure \ref{fig:simulation}, and][]{Price+2018}. 
The disc around AB~Aur does not appear to have a substantial eccentricity, since the continuum and free-free emission associated with the central star is close to the projected centre of the dust ring \citep{Tang+2012, Rodriguez+2014, Tang+2017}.

\newc{The structure of the inner spiral arms also depends on the orbit of the companion. This can be readily seen in figure~2 of \citet{Price+2018}. Co-planar stellar companions do not appear to make quasi-symmetric double spiral arms inside the cavity like the ones seen in AB Aur \citep{Ragusa+2017, Thun+2017, Calcino+2019}. Rather, the spiral arms pile up towards the apocentre of the disc \citep[e.g. see figure 1 of][]{Calcino+2019}.} For these reasons, we expect the binary, should it exist, will be misaligned respect to the circumbinary disc. %\newc{The spirals beyond 0.’'5 \citep{Grady+1999,Fukagawa+2004} are likely due to \textit{cloudlet} capture, which results in arc-shaped reflection nebulosities in the disc as suggested by \cite{Dullemond+2019}. We recall that in this work we do not model the complex geometry of the disc outer regions.}

Previous works on circumbinary disc dynamics around eccentric binaries show that close to polar configurations are stable for inclined discs when the line of nodes of the binary ($\Omega$) is $\sim 90\degree$ \citep{Aly+2015, Martin&Lubow2017, Zanazzi&Lai2018, Cuello&Giuppone2019}. More specifically, a polar configuration is stable if the binary eccentricity vector and the disc angular momentum vector are parallel \citep[see figure~1 in][]{Aly+2015}. We found that placing a companion on the stable polar alignment does not produce a double spiral arm feature inside the cavity of AB~Aur. Instead, this feature is much easier to match when setting the binary eccentricity vector orthogonal with respect to the disc angular momentum vector ($i=90\degree$, $\Omega=0\degree$, $\omega=0\degree$). Even though this configuration is not stable for long-term evolution, it has been shown that misaligned circumbinary discs around an eccentric binary can remain in unstable configurations for many thousands of binary orbits \citep{Smallwood+2019}, where the disc oscillates around the stable configuration. Furthermore, cluster-level star formation simulations indicate that a substantial portion of binary stars will form with circumbinary discs that are highly misaligned \citep{Bate2018, Wurster+2019}. It is also possible for post-formation inflows onto the circumbinary disc to perturb its angular momentum vector to remain misaligned \citep{Dullemond+2019,Kuffmeier+2020}.

\cite{Poblete+2019} showed that a binary companion inclined with respect to the circumbinary disc produces dust asymmetries resembling those seen in AB~Aur. The case with the binary inclined at $i_{\rm B} = 90 \degree$ does not produce dusty clumps such as those observed in AB~Aur. \newc{Their figure~9 shows that the best agreement with the observations at 1.3~mm is found for $i_{\rm B} = 60 \degree$}. On the other hand, in the present work, we show that \newc{a binary with an orbital plane} at $i_{\rm B} = 90 \degree$ produces an excellent match to the spiral morphology seen in $^{12}$CO line emission in AB~Aur. We therefore suggest that there is an inclined binary companion in AB~Aur, with an inclination ranging between $60\degree$ and $90\degree$ \newc{(with $\Omega=0\degree$, $\omega=0\degree$)}. We note that this corresponds to an unstable configuration. \newc{However, such a configuration can still be relevant for young systems or systems experiencing inflows, as discussed above.}

\subsection{Observational Signatures}

In \cite{Tang+2017} the spiral arms have \newc{an apparent} contrast ratio $\sim$~4:1 with the surrounding material. The true contrast ratio could be \newc{as low as 2:1} since the $uv$-coverage is sparse, and more diffuse emission is filtered out. The contrast implies that the spiral arms are either a density or temperature enhancement, or a combination of these. \citet{Tang+2017} suggest the spiral structure they observe maybe be generated by planetary companions; however, it is not clear how feasible it is for planetary-mass companions to generate such a contrast ratio in CO emission. Given the large density enhancements caused by the companion, our inner binary scenario has no difficulty in explaining the observed contrast ratio between the spiral arms and surrounding material. \newc{We note that varying the companion's mass in our simulation changes the contrast ratio between the spirals and the surrounding material. Although this has not yet been thoroughly studied for misaligned circumbinary discs, our results are in agreement with \cite{Bae2018} where it is shown that the higher the companion's mass the higher the density contrast.}

\newc{The best way to distinguish binary-induced spirals in CO emission from planet-induced spirals may not be to study the spirals' brightness alone, but also trace their position closer to the cavity edge.} The pitch angle of outer spiral arms generated by planetary mass companions is small \citep{Rafikov+2002}, and the contrast ratio of the spiral with the background material decreases along the spiral arm as the shock from the spiral weakens \citep[e.g.][]{Zhu+2015, Dong+2015b, Bae2018}. Therefore if the spiral arms are generated by planetary mass companions inside the cavity of AB~Aur, the pitch angle of the spiral arms should be low and their contrast with the background disc material should weaken along the spiral. In contrast, our stellar mass companion creates spiral shocks with a strong contrast ratio far from its orbital position, as evident in the left panel of Figure~\ref{fig:simulation}. We can also see in Figures~\ref{fig:simulation} and \ref{fig:spiral_evolution} that the spiral arms propagate towards the cavity edge, uninterrupted (except when the secondary star catches the arm), and with a higher pitch angle than what is expected for a planetary mass companion.

In Figure~\ref{fig:spiral_evolution}, our closest match to the spiral arms occurs when $t \sim 0.7 \ T_\mathrm{b}$. Going through the panels, we see that the spiral arms propagate towards the cavity edge, uninterrupted, \newc{and without connecting to any outer planetary-mass companions}. Indeed, we \newc{can} see this in our radiative transfer models of \newc{the} simulation; however, the outer part of the spirals is filtered out when we recreate the observing parameters from \citet{Tang+2017}. Therefore, we predict that with better $uv$-coverage the spiral structure should be observed to connect with the cavity.

\begin{figure}
\centering
\begin{center}
    \includegraphics[width=0.45\textwidth]{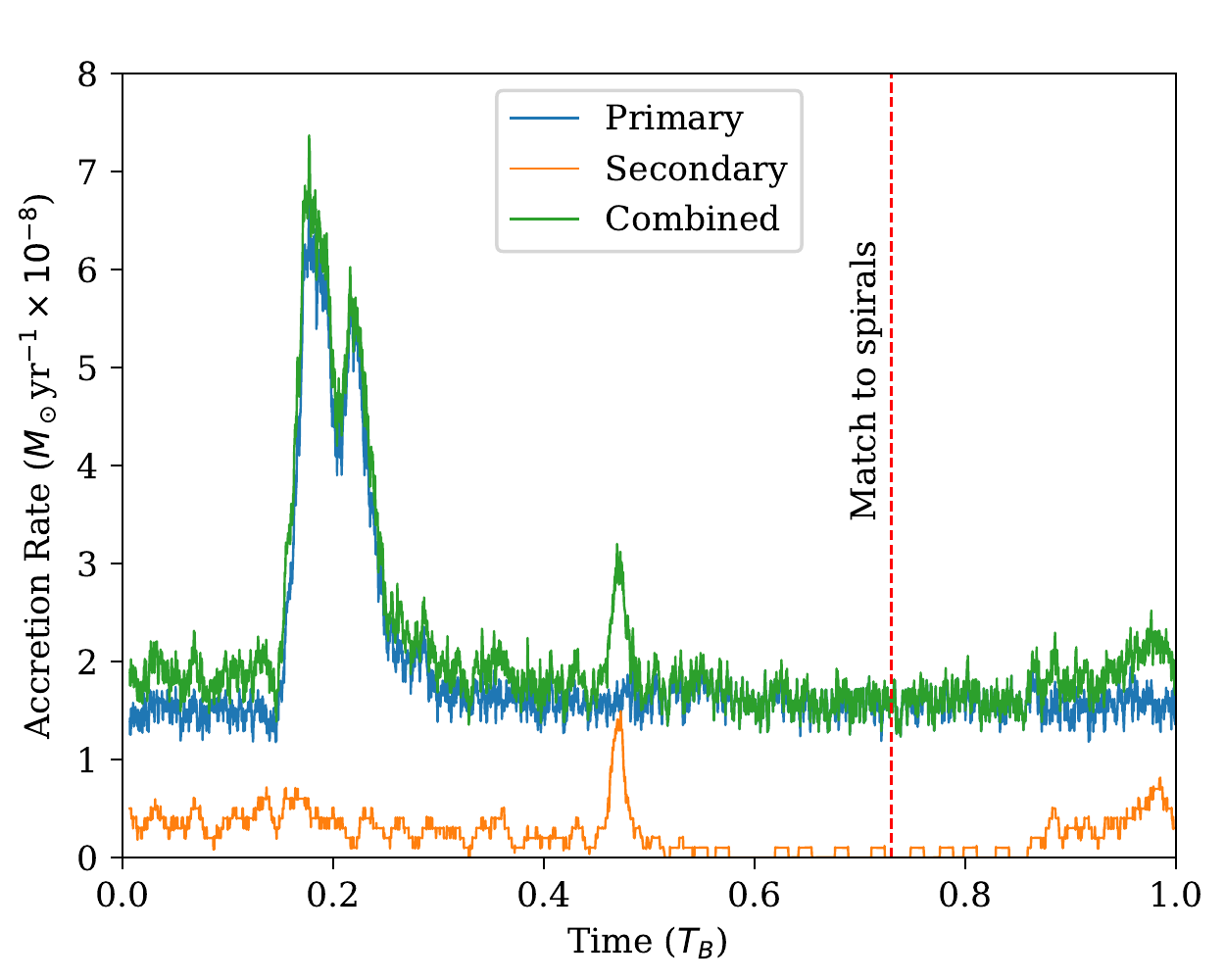}
    \vspace{-2em}
    \caption{The accretion rate of the primary, secondary, and primary+secondary in our simulation during one orbit of the binary. Accretion rate is expressed in $M_\odot$yr$^{-1}$, and time is scaled to the orbital period of the companion. We see a sharp spike in the accretion rate of the primary at roughly $0.2\ T_B$, which is consistent with the current accretion rate of 1.3~$\times 10^{-7}\ M_{\odot}\ \rm yr^{-1}$ seen in AB~Aur \citep{Garcia-Lopez+2006, Salyk+2013}. However the accretion rate is low when we best match the spiral structure. The accretion rate of the companion remains low for almost all of its orbit, except when it intercepts the plane of the disk at roughly $0.5\ T_B$ and $1\ T_B$. }
    \label{fig:accretion}
\end{center}
\end{figure}

The binary scenario could also explain the high accretion rate observed in the system. AB~Aur shows an accretion rate of $1.3 \cdot 10^{-7}\ M_{\odot}\ \rm yr^{-1}$; which is considered high for a star of its kind \citep{Garcia-Lopez+2006, Salyk+2013}. In Figure~\ref{fig:accretion}, we show the accretion rate of the primary star during one orbit of the companion in our simulation. We see that during most of the binary orbit, the accretion rate of the primary remains at roughly $\sim 1.5\cdot 10^{-8}\ M_{\odot}\ \rm yr^{-1}$ However there is a portion of the binary orbit where the accretion rate spikes close to $\sim 7~\cdot 10^{-8}\ M_{\odot}\ \rm yr^{-1}$. This occurs at roughly $t=0.2\ T_\mathrm{b}$, and not at the time when we see the best match to the spiral structure ($t \sim 0.7\ T_\mathrm{b}$, indicated by the red line). Further exploring the companion's parameter space may help to solve this discrepancy. However, we should clarify that our accretion rate simply measures the flux of SPH particles that enter the accretion radius of 1 au for our primary star. We do not model magnetospheric accretion onto the star, or the dynamics of the clumps of gas that enter inside the accretion radius. \newc{Furthermore,} the spike in the accretion rate also varies between orbits, by up to a factor of a few higher than in Figure \ref{fig:accretion}.

%%-------------------------
	
\section{Conclusion}
\label{sec:conclusion}

We performed a 3D SPH gas simulation of a circumbinary disc along with radiative transfer models matching the parameters of AB~Aur. Our results support the hypothesis that there is an unseen stellar companion inside the disc cavity. We conclude that:
\begin{enumerate}
\item The spiral arms in AB~Aur can be explained by the presence of an inner stellar binary that affects the gas within the disc cavity.

\item To reproduce the disc morphology, the proposed inner binary should be of unequal-mass ($q\sim0.25$) and eccentric ($e_{\rm B}\sim0.5$). \newc{More importantly, given our results and the dust morphologies in \citet{Poblete+2019}, we constrain the inclination of the binary with respect to the circumbinary disc to range between $60\degree$ and $90\degree$.}

\item Since the morphology of the inner spiral arms depends on the companion location, our model predicts the location of the companion at $r \sim 0.18\arcsec$ from the central source, with a $\rm PA\sim 100\degree$. 

\item Should our proposed companion be detected, we expect it to have a high radial velocity moving away from the central star. Our model suggest a motion of 6 mas/yr away from the primary star, but this value is model dependent.

\item Motion of the spirals should also be detectable with observations taken on a long enough baseline. However unlike companions on co-planar orbits, we expect the spiral structure to vary with time, and not co-rotate with the companion.

\end{enumerate}

Our present work adds to the growing body of evidence that suggests the spiral arms and dust asymmetries observed in many TDs are not necessarily sign-posts for planet formation. Rather, this class of discs may in fact be made of two distinct sub-populations: genuine planet hosting TDs such as PDS70 \citep{Keppler+2018, Muller+2018}, and circumbinary discs such as HD~142527 \citep{Biller+2012,Lacour+2016,Price+2018}. Future observations will confirm or rule out the circumbinary nature of the disc around AB~Aur.

\section*{Note added in proof}

While this paper was under review, a new study by \citet{Boccaletti+2020} presented VLT/SPHERE observations of AB~Aur, which they interpret as evidence for a planetary-mass companion. Our work suggests a stellar-mass companion is required, but a detailed interpretation to their data in our framework is deferred to follow-up work.

%%--------------------

\section*{Acknowledgements}

We thank the referee Ya-Wen Tang for useful comments and suggestions. We also thank Valentin Christiaens for discussions. PP and JCu acknowledge support from Iniciativa Cient\'ifica Milenio via the N\'ucleo Milenio de Formaci\'on Planetaria. PP, NC and JCu acknowledge support from CONICYT project Basal AFB-170002. JC acknowledges support from an Australian Government Research Training Program Scholarship. DJP and CP are grateful for Australian Research Council funding via DP180104235, FT130100034 and FT170100040. This project has received funding from the European Union's Horizon 2020 research and innovation programme under the Marie Sk\l{}odowska-Curie grant agreements Nº 210021 and Nº 823823 (DUSTBUSTERS). The Geryon2 cluster housed at the Centro de Astro-Ingenier\'ia UC was used for the calculations performed in this paper. The BASAL PFB-06 CATA, Anillo ACT-86, FONDEQUIP AIC-57, and QUIMAL 130008 provided funding for several improvements to the Geryon/Geryon2 cluster. Other calculations were also performed on the \texttt{getafix} cluster hosted by the School of Mathematics and Physics at the University of Queensland, and the OzSTAR national facility at Swinburne University of Technology. OzSTAR is funded by Swinburne University of Technology and the National Collaborative Research Infrastructure Strategy (NCRIS). This work has been partially supported by the Deutsche Forschungsgemeinschaft (grant LO 1715/2-1, within Research Unit FOR 2285 ``Debris Disks in Planetary System''.

\bibliographystyle{mnras}
\bibliography{AB-Aur}

%%--------------------
\appendix
%%--------------------

\section{CO maps for a gas-to-dust ratio of 10}\label{A1}

\begin{figure*}%
\begin{center}

    {\includegraphics[width=0.77\textwidth]{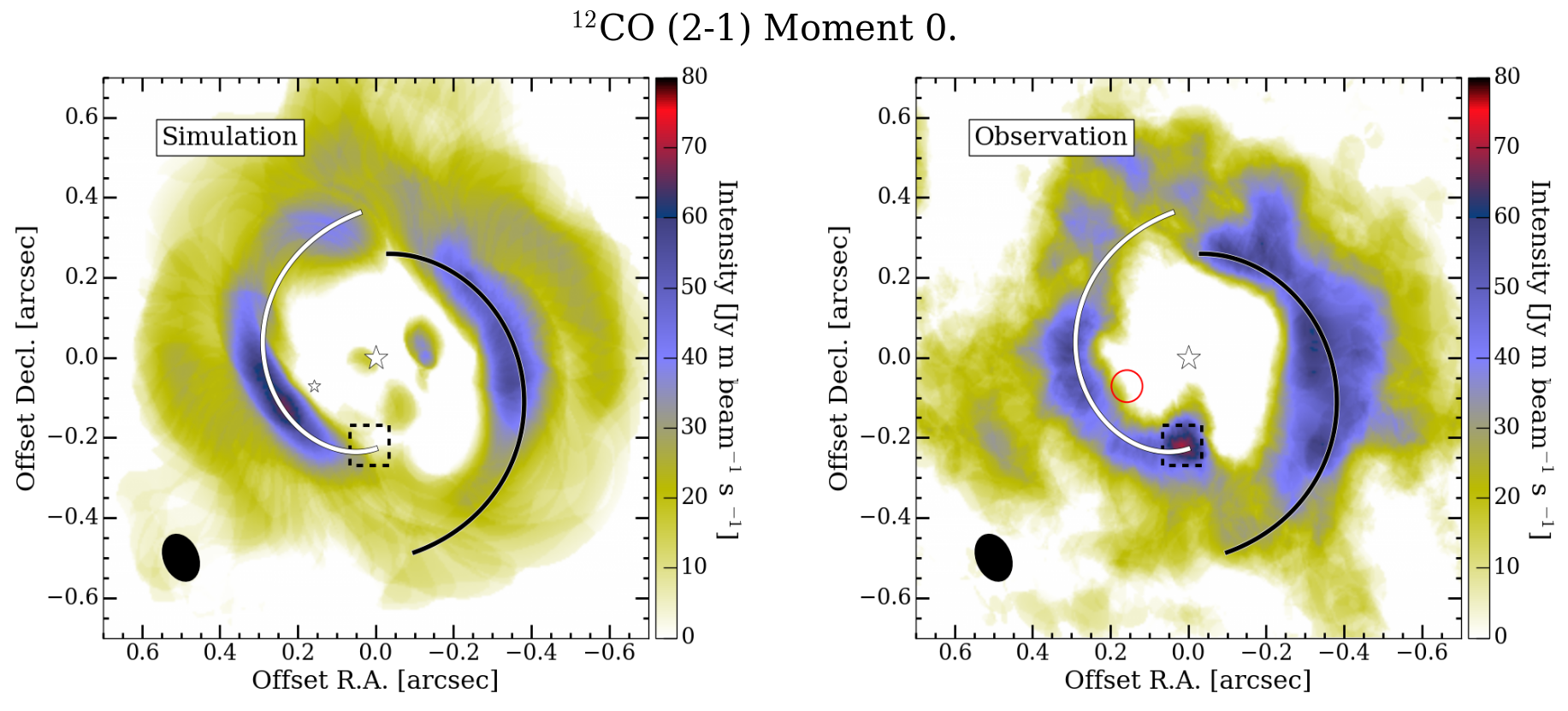}}\\
    {\includegraphics[width=0.77\textwidth]{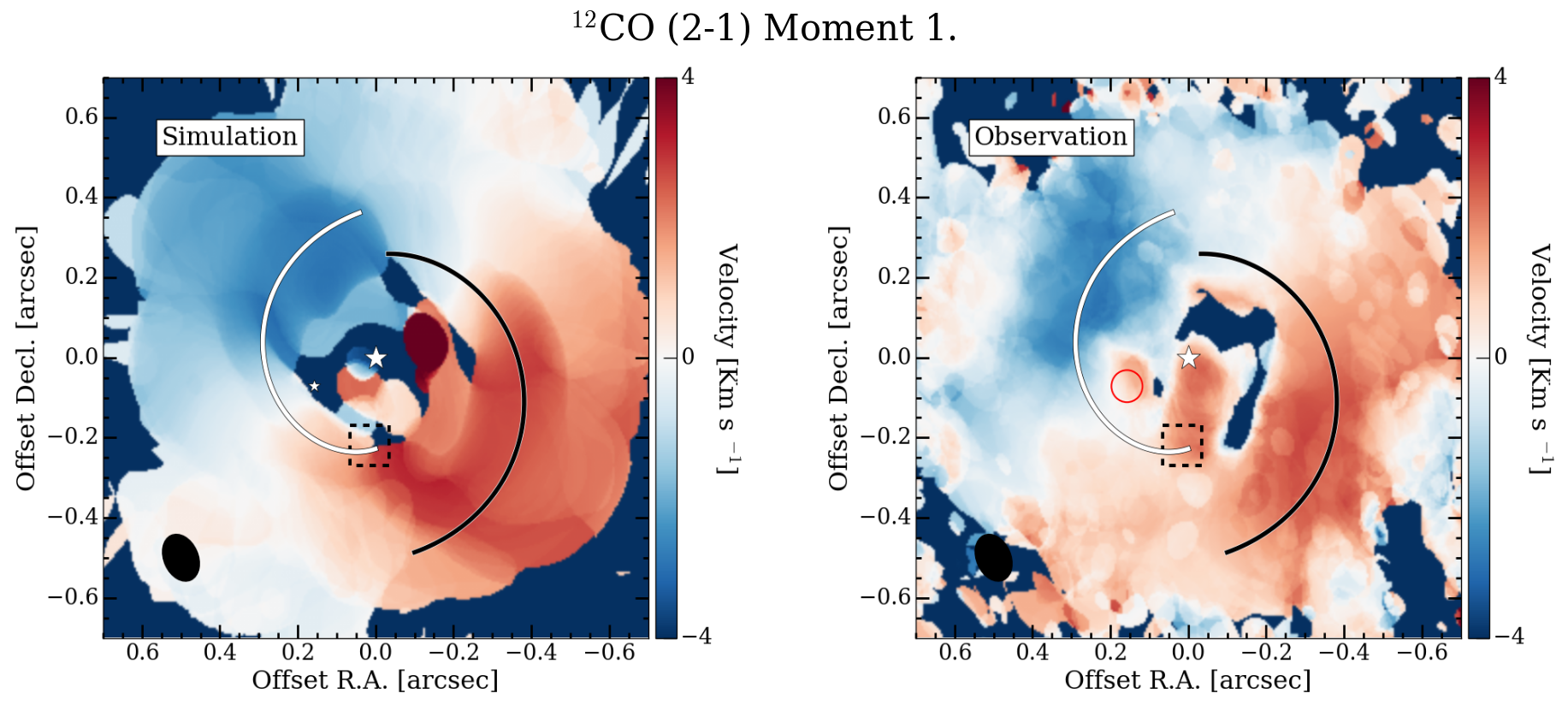}}%
    \caption{Same as Figure \ref{fig:moment_0} but for a gas-to-dust ratio of 10.}
    \label{fig:moment_0_100}%
\end{center}
\end{figure*}

\begin{figure*}
\centering
\begin{center}
    {\includegraphics[width=0.48\textwidth]{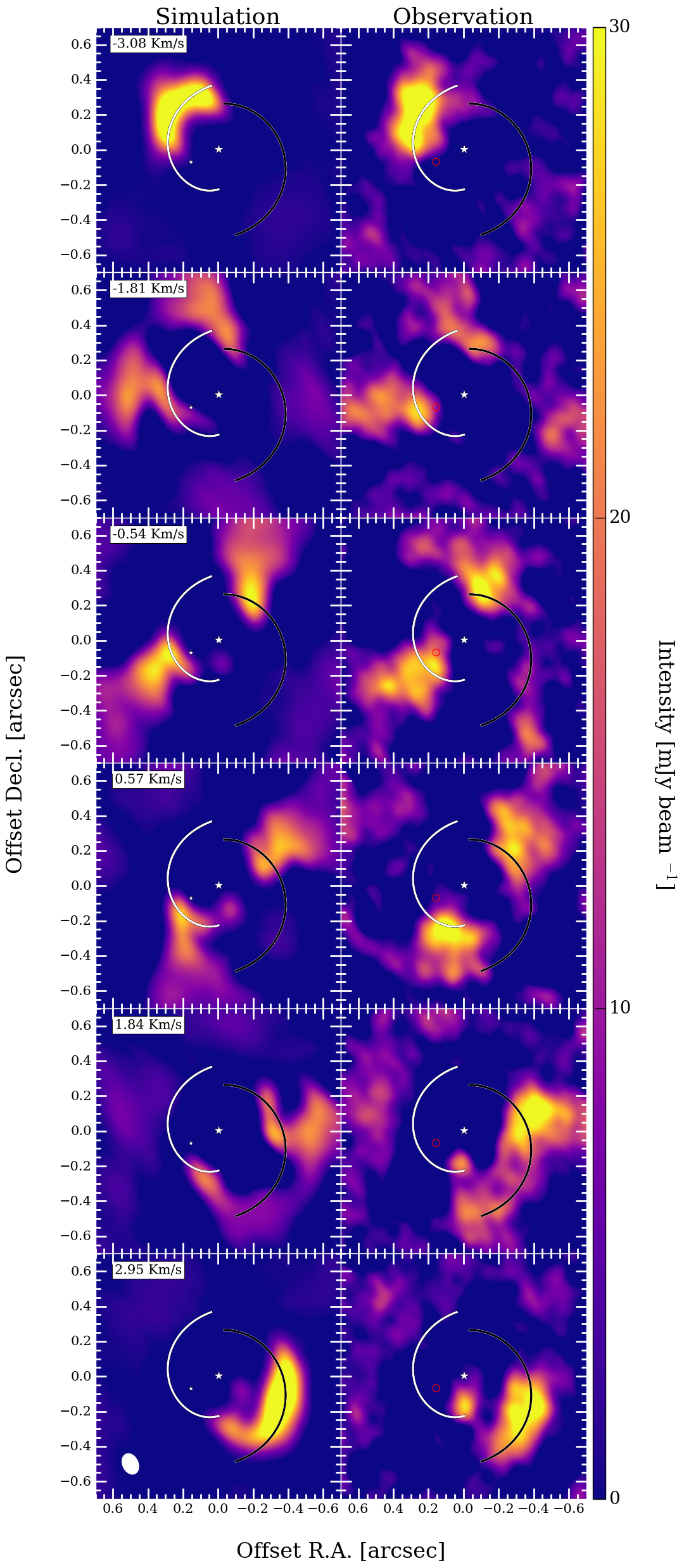}}%
    \caption{Same as Figure \ref{fig:chan_map} but for a gas-to-dust ratio of 10.}
    \label{fig:chan_map_100}
\end{center}
\end{figure*}

\newc{In order to explore the effects of a lower gas-to-dust ratio on the excess of CO emission in the central sector, we computed the same radiative transfer model presented in Section \ref{sec:RT_model} but for a gas-to-dust ratio of 10. The results are shown in Figures \ref{fig:moment_0_100} and \ref{fig:chan_map_100}. With this setup, the models resemble the observed CO emission of the inner regions more closely. This suggests that in the inner regions the gas-to-dust ratio could reach values as low as 10.}

\label{lastpage}
\end{document}